# Duplication of Windows Services

Zhiyong Shan Xin Wang Tzi-cker Chiueh Rajiv Bagai

**Abstract**—OS-level virtualization techniques virtualize system resources at the system call interface, has the distinct advantage of smaller run-time resource requirements as compared to HAL-level virtualization techniques, and thus forms an important building block for virtualizing parallel and distributed applications such as a HPC clusters. Because the Windows operating system puts certain critical functionalities in privileged user-level system service processes, a complete OS-level virtualization solution for the Windows platform requires duplication of such Windows service as Remote Procedure Call Server Service (RPCSS). As many implementation details of the Windows system services are proprietary, duplicating Windows system services becomes the key technical challenge for virtualizing the Windows platform at the OS level. Moreover, as a core component of cloud computing, IIS web server-related services need to be duplicated in containers (i.e., OS-level virtual machines), but so far there is no such scheme. In this paper, we thoroughly identify all issues that affect service duplication, and then propose the first known methodology to systematically duplicate both system and ordinary Windows services. Our experiments show that the methodology can duplicate a set of system and ordinary services on different versions of Windows OS.

---

## 1 Introduction

Virtual machine (VM) is a software layer for constructing multiple execution environments on the host machine. The virtualization can take place at several different levels of abstractions [23][6], including the Instruction Set Architecture (ISA), Hardware Abstraction Layer (HAL), operating system level and application level. ISA-level virtualization emulates the entire instruction set architecture of a VM in software. HAL-level virtualization exploits the similarity between the architectures of the guest and host machine, and directly executes certain instructions on the native CPU without emulation. OS-level virtualization partitions the host OS into containers, by redirecting I/O requests, system calls or library function calls. In this paper, we refer to OS-level virtual machine as container that is more commonly used. Each ***container*** provides a complete OS environment but is separated from other containers running on the same host. Application-level virtualization is typically an implementation of a virtual environment that interprets binaries. Multiple levels of virtualization can differ in isolation strength, resource requirement, performance overhead, scalability, and flexibility. In general, when the virtualization layer is closer to the hardware, the created VMs are better isolated from one another and better separated from the host machine, but with more resource requirement and less flexibility. As the virtualization layer is moved up along the machine stack, the performance and scalability of created VMs can be improved.

Operating system level virtualization can not provide isolation as strongly as that of ISA and HAL levels, but it has some distinct advantages when supporting cloud computing, mobile device security protection, HPC clusters, server consolidation, software-defined networks, and the Grid computing [3][4][29][30][5][6][13][23][26][27][19][31][32]. First, each container (i.e., OS-level virtual machine) requires only a small amount of system resources (e.g., memory, disk space) [5][23] and thus can achieve good scalability. This is because a container shares most files and kernel memory with the host environment. Second, containers and the host environment can conveniently synchronize state changes with each other [5][6][26]. Valid state changes in a container can be committed to the host environment [33], while new updates or configuration of the host environment can be synchronized immediately in a container. Third, tests in publications [27][19][28] show that containers are more efficient than hardware-level virtual machines. Walters et al. [27] systematically evaluates various virtualization technologies for high performance computing in terms of network utilization, SMP performance, file system performance, etc. It shows that the OS-level virtualization provides the best overall performance. HP laboratories [19] evaluated different virtualization technologies for server consolidation and obtained the similar results. Due to these advantages, variants of such OS-level virtualization system are in production use today—e.g., Solaris 10 [14], Virtuozzo [8], and Linux-VServer [22].

A standardized way of implementing OS-level virtualization involves intercepting system calls of the hosting OS, and renaming system resources used in system calls so that the system resources of each container reside in a separate name space. A well-known limitation with OS-level virtualization is that all containers share the underlying hosting kernel and thus

- *Zhiyong Shan is with the Electrical Engineering and Computer Science Department Wichita State University, USA. E-mail: zhiyong.shan@wichita.edu.*
- *Xin Wang is with the Electrical and Computer Engineering Department, Stony Brook University, USA. E-mail: xwan@ece.sunysb.edu.*
- *Tzi-cker Chiueh is with the Computer Science Department, Stony Brook University, USA. E-mail: chiueh@ cs.sunysb.edu.*
- *Rajiv Bagai is with the Electrical Engineering and Computer Science Department*

its states. On the Windows platform, this problem is exacerbated because Windows uses a set of user-level system services, which behave like daemons in a Unix-style OS, to augment the kernel to provide various functions critical to other user-level services and application processes. For example, Windows inter-process communication mechanisms such as COM, DCOM, and RPC, are supported by the RPCSS service (Remote Procedure Call Server Service). If the RPCSS is shared among containers, it is impossible to execute two instances of the same application, e.g., MS Office Assistant, in two distinct containers. It remains to be an important problem to duplicate Windows system services such as RPCSS so that the large installed base of Windows applications can concurrently run multiple instances in different containers without interfering with one another. The goal of this research is to solve this problem. As sharing services among different containers affects the isolation and security of containers, duplicating services into different containers can reduce the sharing and thus significantly improve container isolation and security.

Conceptually, duplicating Windows system services means replicating a distinct instance of each Windows system service in every container. However, it is non-trivial to completely achieve this goal for the following reasons. First, some Windows system services, e.g., RPCSS, are designed to prevent creating, registering or running multiple instances of the same system service simultaneously on top of a Windows OS. Second, to duplicate Windows services, one needs to intercept and modify inter-process communications, inter-service cooperation, registry key manipulations, etc. However, because the details of these interactions are largely proprietary, it is difficult to make necessary modifications to them. Third, in some cases, it is not sufficient to just correctly handle these interactions. Some Windows services hard-code certain system resource names in their binaries, and they will not function properly when run inside containers which often have their system resources transparently renamed.

Through an extensive effort of investigating the design principles of Windows services and with substantial trial-and-error, we have developed the first methodology for duplicating Windows services. To evaluate the effectiveness of this methodology, we apply it to several critical system services, including RPCSS, on multiple versions of Windows OS. All of the critical system services are successfully duplicated. To the best of our knowledge, this is the first paper that demonstrates the success of running multiple instances of critical Windows services such as RPCSS concurrently on top of a Windows OS. Moreover, the same methodology allows multiple instances of ordinary Windows services such as IIS, Apache or Mysql to run simultaneously on a Windows machine, which are often the core components of cloud computing [24][25].

Other types of OS, e.g., Linux and Mac, have similar requirements and challenges on duplicating critical system applications like Windows services. For example, they often need inter-process communications across container boundaries and involve hard-coded resource names. Although our solutions are tested on the Windows platform, our methodology is useful when addressing the similar issues on other types of OS platform. Due to the large efforts required for OS kernel level prototyping, in this paper, we only demonstrate our solutions on two different Windows platforms.

The contributions of this paper are three-fold:

1. We thoroughly identify all issues concerning the service duplication.

2. We propose the first known solution to the problem of duplicating Windows services---the key technical obstacle to container-based virtualization on Windows.

3. We design a novel mechanism to automatically enable necessary communications across container boundaries. Meanwhile, we enforce access control to reduce the impact on isolation among containers to the least degree.

## 2 BACKGROUND

### 2.1 Windows Service Architecture

Like daemons in a UNIX-style OS, Windows services run as background processes, rather than being directly controlled by an interactive user. However, compared to a UNIX daemon, Windows services have more complicated software architecture and implementation. The architectural overview of Windows services is shown in Figure 1. We observe that, there are two classes of Windows services: Windows system services and Windows Ordinary services. ***Windows system services***, such as RPCSS, offer functions essential to many user-level services and applications across the system. ***Windows ordinary services***, such as IIS and Apache, only perform specific application tasks while not providing functionality to other services. Both types of Windows services are not bound with the kernel and thus can be duplicated into containers.

***Windows core processes*** in Figure 1 refer to system processes but not a type of Windows services. Windows core processes are extensions of the kernel and provide basic functionalities for the start-up and running of Windows services, and include such processes as Service Control Manager (SCM) and Winlogon. Unlike Windows services, core processes are bound with the kernel and should not be duplicated in containers. Specifically, there are two reasons. First, a core process might be responsible for preparing an essential running environment of the OS, for example, SMSS creates Win32 subsystem. Creating multiple instances of SMSS will fail, because Windows OS kernel can not support multiple instances of Win32 subsystems. Second, a core process might fulfill a set of OS API functions which extend those of the OS kernel. For example, SCM process fulfills service management API functions. Duplicating such form of core processes would be disabled by the kernel, which was observed from our experiments. Consequently, core processes like SCM and SMSS can not be duplicated and thus must stay in the host environment.

Windows services are started in a way very different from normal user-level processes. All Windows services are managed by a core process called the *Service Control Manager* (SCM), which maintains an internal database that records the group of registry keys, the name, the image path, and the start-up parameters for each Windows service. A set of predefined operations must be called in an orderly fashion, as shown in Figure 1, so as to create, initiate, and run a Windows service under SCM's supervision. Step 1 is to create a new Windows service by calling the API function CreateService(), in which the SCM adds the service's name, image path, and start-up parameters into the SCM database. Step 2 is to start a service process by invoking the API function StartService(), in which the SCM spawns a service process according to the service's information in the SCM database. In Step 3, a service process that just gets started must immediately tell the SCM what services it offers by sending the SCM a service table. In Step 4 the newly started service process registers its service name with the SCM. If a service name is already registered in the SCM or is not in any service table, the SCM denies the service name registration request. By denying multiple requests to register the same service, the SCM prevents multiple instances of the same service from running. In Step 5, the service process reports to the SCM when it is ready to accept service requests. In Step 6, when a user process requests access to a target service, the SCM forwards the request to the corresponding service. Finally, during the startup time or runtime, a service process may have to interact with other services and Windows core processes to, for example, login, authenticate itself, or launch its COM server. A Windows service process may hang or act unexpectedly if any of the above interactions fails in some way.

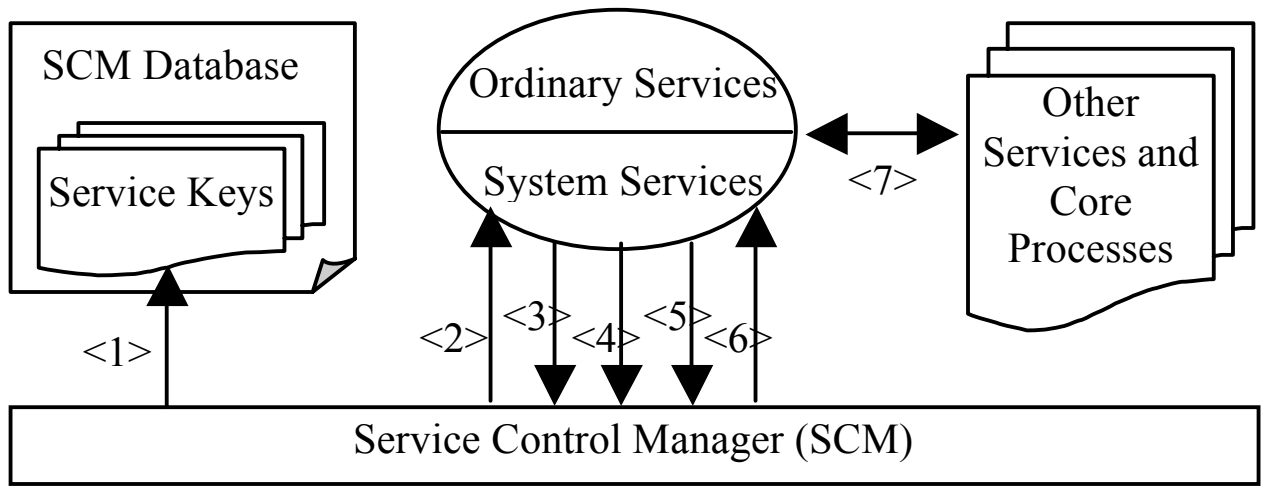

Fig. 1. Software architecture for Windows services

Communications between Windows services or between Windows services and Windows core processes are through complex and often undocumented inter-process communication (IPC) mechanisms. Compared to the UNIX-style OS, the Windows OS supports a wider variety of IPC mechanisms, including mutex, event, timer, semaphore, shared memory, mailslot, pipe, socket, RPC, LPC, DDE, COM, Windows message, data copy, and clipboard. Without a full understanding of how Windows services interact with one another and with Windows core processes, it is difficult to duplicate these services by properly restricting or transforming these interactions.

Some Windows services are in the form of a DLL (Dynamically Linked Library) and their run-time incarnation is different from those Windows services that are processes instantiated from an EXE file. DLL-based services require additional information stored in the SCM database and different startup parameters when compared with process-based services. These additional complexities further complicate the task of duplicating DLL-based Windows services.

Finally, because most Windows services are close-source, their internal implementation details are rarely documented in any open literature, which include the internal logic, inter-process communication, inter-service cooperation, related registry entries, etc. Consequently, it takes a great deal of effort to investigate and reveal the implementation details of Windows services to the extent that the resulting knowledge enables us to duplicate these services. In this project, we used three techniques to understand the design principles of Windows services. First, we traced the kernel-level and Windows API-level calls that a service process invokes at run time in order to determine the set of resources a service process accesses, and the inter-process interactions the service process is engaged in. Second, we used the tool ProcessExplorer[7] to find out the inter-process communication objects the service process uses to talk with other Windows services. Lastly, we disassembled a service's binary code to uncover any hard-coded service names in a service's binary code and API function calls that use them.

## 2.2 FVM

To implement Windows service duplication, we choose *FVM* (*Feather-Weight Virtual Machine*) [5], a Windows-based OS level virtualization technology, as the starting point. A major design goal of FVM is to allow efficient resource sharing among containers so as to minimize container start-up/shut-down cost and scale to a large number of concurrent container instances. As a result, FVM serves as a great platform for developing failure-tolerant or intrusion-tolerant applications, for instance, "scalable web site testing" [6], which isolates the underlying host machine from the potential malignant side effects of browser attacks launched from un-trusted web sites.

In FVM, the host is a Windows machine on which normal user processes run, and these user processes are grouped into containers that are isolated from one another. The key idea behind FVM is *namespace virtualization*, which renames system resources through a virtualization layer that sits below the OS system call interface. The Windows OS uses different name spaces for different system resources, such as files, registries, kernel objects, network addresses, Windows services, and window classes. The FVM layer manipulates the names of all these resources when a process makes system calls to access them. Through resource renaming, the namespaces visible to processes in one container are guaranteed to be disjoint from those visible to processes in another container. As a result, two containers never share any resources and therefore cannot interact with each other directly. For example, suppose an application in one container (say vm1) tries to access a file /a/b, then the FVM layer will redirect it to access /vm1/a/b. When a process in another container (say vm2) accesses /a/b, it will try a different file, i.e., /vm2/a/b, which is different from the file /a/b in vm1.

However, to avoid unnecessary duplication of common system resources that leads to significant start-up and/or run-time overhead as exhibited by many heavyweight virtual machine technologies, FVM enables containers to share most resources with the host by default while isolating state changes attributed to each individual container through a special copy-on-write scheme. That is, a newly created container initially shares all the resources in the host. If processes in the container make only read accesses to the system resources, they can simply access the shared resources in the host, and the container does not own any private resources, until processes in the container try to modify a shared system resource, at which point FVM copies the modified system resource and forks it into different versions. Therefore, the resource requirement of each container is significantly lower under FVM than that with HAL-level virtualization technologies such as VMware [5].

Despite the fact that FVM supports renaming for most name spaces on the Windows platform, the degree of isolation among containers that it provides is still not as strong as HAL-level virtualization. Previous versions of FVM give a few exceptions to the communications between Windows system services in the host and applications in containers because system services can not be duplicated into each container. This might affect the isolation capability and system security. To improve the isolation and security, in this work, we explore techniques to duplicate system services into containers.

## 3 Windows Service Duplication

### 3.1 Problem Statement

The ultimate goal of Windows service duplication is to replicate a new instance of an arbitrary Windows service and assign the instance to every new container that spawns the Windows service. To implement Windows service duplication on top of FVM, we need to address the following four main design issues.

First, it is impossible to call standard API functions to create a DLL-based service, such as the RPCSS service, because a DLL-based service usually runs as a thread inside a service host process called *svchost*. Moreover, aside from common SCM database information fields shared by all Windows services, a DLL-based service needs to insert special SCM database entries under the registry keys owned by its corresponding svchost process. Unfortunately, it is not easy to use API functions to create these registry entries at the time of creating a DLL-based service. DLL-based services also require special SCM database fields to store the binary image path and start-up parameters of their service host processes. In this research, we have developed a methodology that can duplicate both EXE-based and DLL-based services in a uniform way.

Second, because the SCM process is a user-level Windows core process that has been tightly tied with the operating system, it cannot be duplicated and always runs in the host. Meanwhile, because all Windows services are started by the SCM, it means that all Windows services should also run in the host, according to the original FVM architecture. To be able to assign a Windows service to a container, we need to determine on behalf of which container the SCM process starts a new Windows service, by observing the interactions between containers and the SCM.

Third, complicated inter-service interactions exist between duplicated services and Windows core processes. These interactions are typically embodied through Windows IPC objects, but their exact implementation details are mostly proprietary and thus never documented publicly. Moreover, as presented in Section 2.1, Windows core processes run in the host and cannot be replicated in every container, their interactions with other processes thus must be explicitly allowed to cross container boundaries. Unfortunately, allowing a Windows core process to be shared by multiple instances of the same Window service that run in different containers may cause some unexpected conflicts and failures. Therefore, it is necessary to intercept these communications and properly transform them whenever desirable so that these communications can proceed and the original execution semantics for duplicated services are preserved.

Finally, renaming of services is an important technique used in service duplication. However, some services embed hard-coded service names in their executable binary files. More specifically, because the original developers did not anticipate that a service program may be replicated with multiple service names, they simply hard-coded a fixed service name in the program and used it as an argument in subsequent calls to Win32 API functions. When such service programs run with a different service name than the hard-coded one, it would behave incorrectly. To solve this problem, one needs to first pinpoint the locations of these hard-coded service names in the service binaries through an extensive reverse-engineering effort, and then transparently modify these hard-coded service names.

UNIX-style operating systems such as Linux and Mac have similar issues when duplicating at the system call interface applications that resemble a Windows service, for example, a daemon. Specifically, among the four issues analyzed previously, Linux and Mac have similar issues. Corresponding to the second issue, Mac has core processes similar to SCM on Windows, e.g., launchd, which is responsible for launching daemon processes and are tightly tied with the OS. Regarding the third issue, some processes, e.g., sshd and getty on Linux and launchd on Mac, have to run in the host environment in order to provide services to other applications running inside containers. Hence, it is necessary to allow such interactions while offering strong isolation among containers. As for the last issue, Linux and Mac applications which have hard-coded application names in their executable binaries could break the OS-level virtualization logic. For example, Lynx browser and osgViewer on Linux, as well as Loom Game Engine on Mac have hardcoded application names.

Given the issues are similar, we believe the proposed methodology to Windows service duplication could be useful to OS-level virtualization on Linux and Mac.

### 3.2 Service Duplication Methodology

In this paper, we propose a novel methodology to duplicate Windows services that could address all four issues above. As depicted in Figure 2, it consists of the following four steps:

(1). Logically duplicating a Windows service by creating a separate entry in the SCM database with a new service name.

(2). Physically duplicating a new instance of a service by starting a service process and putting it in its corresponding container.

(3). Managing the inter-service interactions between duplicated services and Windows core processes.

(4). Renaming hard-coded service names embedded in service binaries.

In the rest of this section, we describe these four steps in greater detail.

### 3.3 Logically Duplicating a Windows Service

In order to duplicate a Windows service in a container, we first create a new name for the service in the SCM database, which SCM accesses to start a service. As a result, SCM treats the duplicated service as a different service and does not prevent it from running concurrently with the original one.

There are two ways to duplicate a Windows service. One is for EXE-based services by calling an API function of service creation with a duplicated service name, like “ServiceName-vmX”, where ServiceName is the original name of the service to be duplicated and X is the container ID of the container initiating the service duplication request.

The other is for DLL-based services. The API function of service creation can not create a new instance of a DLL-based service that is identical to the original instance, because it could neither set up a DLL-based service name in the SCM database nor add the DLL-based service into a service host process that adopts a number of DLL-based services as threads. Another method is to modify the SCM database directly in order to add a new DLL-based service. However, this method could not start the new service immediately after the SCM database modification, because SCM cannot incorporate the SCM database modification without rebooting the whole system.

Therefore, we choose to combine these two methods to create a new DLL-based service as follows:

(1) Call the service creation API function with a duplicated service name, like “ServiceName-vmX”, to inform SCM of the creation of a new service. This way we can launch the newly created service without rebooting the entire system, because SCM can record the information of the new service in the memory immediately.

(2) Make a copy of the SCM database entries associated with the duplicated service, and modify three places in the copied SCM database entries. First, the original service name “ServiceName” should be changed to the duplicated name “ServiceName-vmX”. Second, the names of the services on which the duplicated service depends should also be changed from their original service names to duplicated service names. Last, the start type of the new duplicated service should be set to manual start rather than automatic start, because a duplicated service should be started after the boot-up of a container. With these SCM database entries, a duplicated service will behave exactly the same as the original service from which it is cloned.

### 3.4 Physically Duplicating a Windows Service

Creating an instance of a Windows system service in a container needs to inform the SCM to launch a new service process. However, the issue is how to determine which container the newly launched service process should be placed in.

At the time when a container creates a service, we record a FVM flag and the container’s ID in the SCM database entry that contains the service’s binary path and start-up parameter. When the SCM launches a service process, FVM’s in-kernel monitor moves the service process into its corresponding container according to the FVM flag and container ID in the process’ pathname and parameters.

For a DLL-based service, SCM starts it as a thread inside a service host process. In this case we record a FVM flag and a container ID at the end of the service process’ parameter list in the SCM database. For example, the original service A has a parameter like “-k A”, the duplicated service A will have a parameter like “-k A-vmX”, “X” again represents the ID of the container that requests a new instance of service A. When the new service A’s process is launched with parameter “-k A-vmX” and FVM’s in-kernel monitor sees a process with a start-up parameter containing “-vmX”, it adds the process into the container whose container ID is X.

For an EXE-based service, SCM starts it as an independent process. We first copy the EXE file into the container’s workspace and then record its pathname in the SCM database. Because the pathname is in a container, it will naturally contain a FVM flag and a container ID. For example, the pathname of the original service B’s image is “c:\B.exe”, and the pathname of its duplicated version becomes “c:\fvm\VM -X\C\B.exe”, where ”X” represents the ID of the container in which the duplicated service is to be placed. When FVM’s in-kernel monitor sees a process with such an image pathname, it will put the process in the corresponding container.

### 3.5 Maintaining Existing Inter-Service Interactions

As presented in Section 2.1, Windows core processes, e.g., SCM, Lsass, and WinLogon, are closely tied to the Windows OS and cannot be duplicated. Duplication means running a new instance in each container. These core processes therefore run in the host. The duplicated services inside a container can not see them but can interact with them through IPC objects. Under FVM, when a process in a container interacts with other processes in the same container using IPC, FVM’s in-kernel monitor intercepts the IPC requests and renames the resources referred to in the parameters of the requests. However, for the IPC requests used by a duplicated service running in a container to interact with the core processes which run in the host, the arguments used in the IPC requests should not be transformed. For instance, Windows services often connect to the SCM process through a named pipe called NtControlPipe. When a duplicated service running in a container tries to open the named pipe, if the pipe name argument was changed to NtControlPipe-vmX, the duplicated service would not be able to interact with SCM because SCM runs in the host and still uses the original pipe name NtControlPipe.

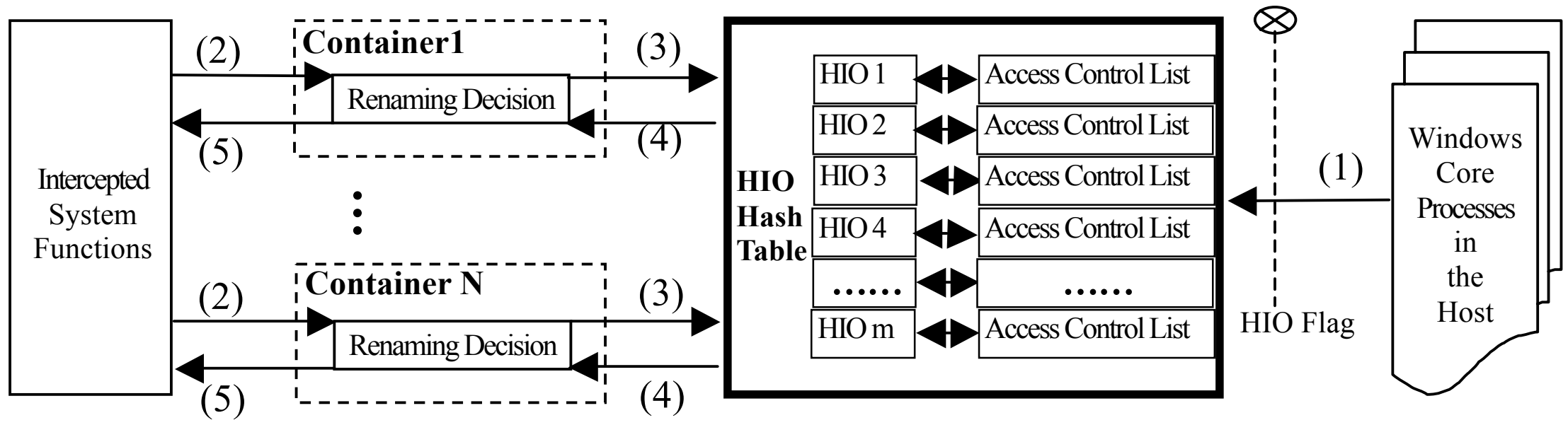

Fig. 3. The two HIO tables used to facilitate interactions between duplicated services and Windows core processes

One way to fix this problem is to identify all IPC calls used by duplicated Windows services to interact with Windows core processes and avoid renaming the resource arguments used in these IPC calls [15]. However, it is very challenging to identify all such IPC calls, because the implementation details of Windows core processes are mostly un-documented. To identify long-term IPC objects and IPC objects used after process start-up, we dynamically analyze the shared IPC objects between Windows core processes and duplicated services using the tool ProcessExplorer. To pinpoint short-term IPC objects and IPC objects used during the process start-up, we logged IPC-related system calls and certain Win32 API function calls during a service's execution in a container and compared the resulting log with that associated with the same service's execution in the host. Through this process, all the IPC objects used by Windows core processes on Windows 7 and 10 have been identified, and they are listed in Table 1. When a Windows service running inside a container utilizes an IPC object, FVM will not rename the IPC object if it is in the list, thus enabling the service to seamlessly interact with the Windows core processes.

However, the above method is non-portable because it is platform- dependent. Once moving to a new Windows version, we have to look for the IPC objects again. In order to address the issue automatically and efficiently, we design a table to help find all such IPC objects, say *Host IPC Objects* (*HIO*), automatically. An HIO can serve for the communications between a duplicated service and a Windows core process. The table stores all the IPC objects of core processes in the host. To reduce the time for searching an HIO from the table, the table is designed as a hash table. When FVM is booted up, the IPC objects are read from core process' space to the table. Every time a Windows service in the container requests an access to an HIO, it queries the table using a hash value of the requested IPC name.

Using HIOs to access core processes in the host means communication across the container boundary, which affects container isolation and security. Hence, we further design an Access Control List for each HIO in the table. Each list contains the names of the Windows services that need to access the HIO, while the ones not on the list can not access the HIO. Thus, the impact on container isolation and security is reduced to the least level.

To automatically build up the access control lists, we further designed a mechanism. That is, every time when a duplicated Windows service requests an access to an HIO object, we insert its name into the HIO object's access control list. We performed a serial of experiments which showed that, after running the system in a secure environment for certain time, the HIO table and access control lists would not be updated any more. This is because that the set of HIO objects and the set of Windows services in a given system are limited and stable. Actually, there are only 20 HIO objects on Windows OSes, which are listed in Table 1. Thus, we can set up a flag, named HIO Flag, to indicate that the table will no longer be updated.

Figure 3 shows our mechanism using the hash table and HIO flag to facilitate interactions between duplicated services and Window core processes in the host. The bracketed numbers in the figure represent the working steps of the mechanism. When the HIO flag is not set, in step (1), all names of the IPC objects created by the core processes in the host are read and stored into the HIO Table; in steps (2) and (5), when a Windows service in a container tries to access an HIO, it sends a request to the Renaming Decision module which is responsible for renaming objects and then waits for the decision result. In steps (3) and (4), the Renaming Decision module checks whether the object is in the HIO Table. If the object exists, it adds the Windows service name into the access control list and returns the original HIO name. When the HIO flag is set, in steps (3) and (4), we further check whether the access control list allows the request but do not update the list.

TABLE 1
IPC OBJECTS INVOLVED IN THE INTERACTIONS BETWEEN DUPLICATED SERVICES AND WINDOWS CORE PROCESSES

| IPC | IPC Objects |
| --- | --- |
| Port | \RPC Control\DNSResolver<br>\RPC Control\ntsvcs |
| Named Pipe | \Device\NamedPipe\net\NtControlPipe* (* represents an arbitrary number)<br>\Device\NamedPipe\svcctl<br>\Device\NamedPipe\ntsvcs<br>\Device\NamedPipe\EVENTLOG<br>\Device\NamedPipe\samr |
| Mutex | \BaseNamedObjects\DBWinMutex<br>\BaseNamedObjects\RasPbFile<br>\BaseNamedObjects\SHIMLIB_LOG_MUTEX<br>\BaseNamedObjects\ShimCacheMutex |
| Section | \BaseNamedObjects__R_ 0000000000da_SMem__<br>\BaseNamedObjects\DBWIN_BUFFER<br>\BaseNamedObjects\ShimSharedMemory |
| Event | \BaseNamedObjects\ScmCreatedEvent<br>\BaseNamedObjects\SvcctrlStartEvent_A3752DX<br>\BaseNamedObjects\crypt32LogoffEvent<br>\BaseNamedObjects\userenv: User Profile setup event<br>\BaseNamedObjects\DINPUTWINMM<br>\SECURITY\LSA_AUTHENTICATION_INITIALIZED |

### 3.6 Renaming Hard-Coded Service Names

After the above three steps, FVM can correctly duplicate many Windows services. However, some Windows services, e.g., RPCSS, still could not be duplicated. By disassembling these services' binary files, we found the root cause lies in the fact that some service binaries include hard-coded service names. Figure 4 shows the disassembly result of several code fragments of the binary file *rpcss.dll*. In these code fragments two hard-coded RPCSS service name strings are used as input arguments to the string processing function RtlInitUnicodeString() and to the service management function OpenServiceW(). Supposed two instances of the RPCSS service are started one after the other in two different containers. Both of them call OpenServiceW() to register themselves to the SCM process in the host. The OpenServiceW() function always uses the hardcoded service name "RPCSS" as a parameter to register to SCM. Thus, the OpenServiceW() call from the second instance will be rejected by SCM because the former instance has already registered the service name "RPCSS" into the SCM database, and consequently the second instance's start-up fails.

```
76A9752F  push      esi
76A97530  push      00000004h
76A97532  push      SWC76A975C4_RpcSS
76A97537  push      [L76ABE8B4]
76A9753D call       [ADVAPI32.dll!OpenServiceW]
……
76A975C4  SWC76A975C4_RpcSS:
76A975C4  unicode 'RpcSS',0000h
……
76A97C0C  SWC76A97C0C_RPCSS:
76A97C0C  Unicode 'RPCSS',0000h
……
76A989FF  mov       edi,[ebp+0Ch]
76A98A02  mov       [L76ABE390],esi
76A98A08  push      [edi]
76A98A0A  mov       esi,[ntdll.dll!RtlInitUnicodeString]
76A98A10  lea       eax,[ebp-18h]
76A98A13  push      eax
76A98A14  mov       [ebp-08h],ebx
76A98A17  call      esi
76A98A19  push      SWC76A97C0C_RPCSS
76A98A1E  lea       eax,[ebp-10h]
76A98A21  push      eax
76A98A22  call      esi
76A98A24  push      00000001h
76A98A26  lea       eax,[ebp-10h]
76A98A29  push      eax
76A98A2A  lea       eax,[ebp-18h]
76A98A2D  push      eax
76A98A2E  call      [ntdll.dll!RtlEqualUnicodeString]
……
76AA8EC4  call      [ADVAPI32.dll!OpenSCManagerW]
76AA8ECA  cmp       eax,edi
76AA8ECC  mov       [ebp-04h],eax
76AA8ECF  jz        L76AA8F16
76AA8ED1  push      ebx
76AA8ED2  push      esi
76AA8ED3  push      000F01FFh
76AA8ED8  push      SWC76A97C0C_RPCSS
76AA8EDD  push      eax
76AA8EDE  call      [ADVAPI32.dll!OpenServiceW]
```

Fig. 4. Disassembled code fragments of *rpcss.dll* reveal that two hard-coded RPCSS service name strings are used as input arguments to API functions *RtlInitUnicodeString()* and *OpenServiceW()*.

We analyzed all of the service binaries in different versions of Windows OS. 18.7% of the services have the hard-coded service name problem. Our algorithm designed to identify the problem has four steps: (1) for a given binary, querying the corresponding service name from registry; (2) disassembling the binary; (3) searching the service name in the disassembly code to find hard-coded service names; (4) checking call instructions to see whether the hard-coded names are used as arguments of a function. The results show that most hard-coded service names are used by service related API functions. The reason is that we are checking hard-coded "service" name. The rest of the hard-coded names are used by string manipulation functions. This is also understandable because the hard-coded names are strings.

Accordingly, we intercept service related and string manipulation functions issued by a duplicated service, and check if the associated argument is a duplicated service name or an unmodified service name. If it is an original service name, we change it to the corresponding duplicated service name, because functions in duplicated service processes should only use duplicated service names. With this fix, we were able to resolve the "hard-coded service name" problem.

## 4 Evaluation

To demonstrate the effectiveness of the proposed Windows service duplication scheme, we developed a prototype based on FVM[6]. To verify the feasibility of our scheme on newer versions of Windows OS, we performed a serial of experiments on Windows 10. The results show that all critical points of our scheme can be implemented.

The Windows service duplication prototype consists of two parts. The first part is a new module in FVM's user-level management tool. It duplicates a Windows service by creating a new entry in the SCM database and starts the target service process according to the user's requirement. The second part involves modifications to FVM's virtualization layer in the kernel and systems library, puts a new instance of a duplicated service into its corresponding container, enables necessary inter-service interactions and renames hard-coded names in the service binaries. With this prototype, we evaluated its functional correctness and performance overhead, and the results are detailed in the next two subsections.

### 4.1 Functionality Evaluation

To verify the functionality of the prototype, we apply it to duplicate several important Windows services including the RPCSS and IIS service group, as well as more services

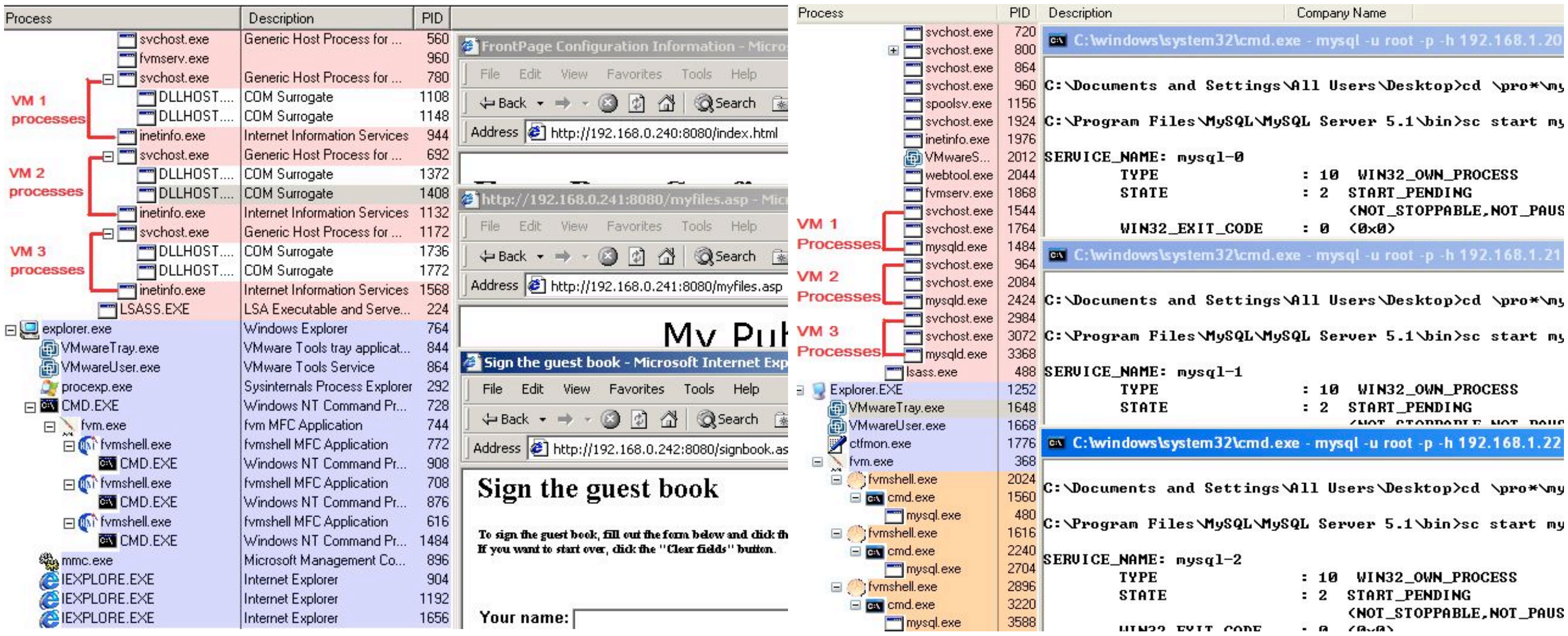

Fig. 5. Three instances of the IIS web server with the same name inetinfo.exe are running on top of a single Windows OS, each of which resides in a separate container and is shown to supply a page to a web browser.

Fig. 6. Three instances of the MySQL database server with the same name mysqld.exe are running on top of a single Windows OS, each of which resides in a separate container and is interacting with a distinct login client.

on different versions of Windows. We present the details as follows:

*4.1.1. Duplicating RPCSS And DcomLaunch*

RPCSS is a fundamental service on the Windows platform that provides RPC/COM/DCOM functions to other Windows services and applications. There are about 100 Windows services running on Windows 10 and nearly half of them depend on RPCSS. Therefore, duplication of RPCSS is essential to OS-level virtualization on Windows. On Windows 10, the functions of RPCSS are split into two services, DcomLaunch and RPCSS. These two services co-operate with each other to provide RPC/COM/DCOM functions to applications. Using the prototype, we duplicated both of them.

To validate that the duplicated RPCSS and DcomLaunch work we performed a serial of experiments inside containers: (1) observing whether multiple duplicated RPCSS can boot up and create all IPC objects correctly; (2) writing a test COM server and running it to retrieve Windows version information; (3) starting Microsoft Office Assistant and testing its functions; (4) opening Excel inside a Word document and testing its functions; (5) starting multiple instances of IIS (Internet Information Services) web servers and testing the functions; (6) launching more services that are presented in Section 4.1.3; (7) starting 24 types of more services. All these experiments succeeded. As services or applications in (2) ~ (7) rely on the functions of RPCSS and DcomLaunch, we thus conclude that the duplicated RPCSS and DcomLaunch work correctly.

*4.1.2. Duplicating IIS*

IIS (Internet Information Services) service group consists of five Internet-based services including W3SVC service for WWW, MSFTPSVC service for FTP, SMTPSVC service for SMTP, NNTPSVC service for NNTP, as well as IISADMIN service for the management of IIS. IIS is the world's second most popular web server in terms of the number of websites using it. Using the prototype, we duplicated all the services in the IIS service group, including W3SVC.

Figure 5 shows a screen shot of running three IIS web server instances in parallel on a single Windows machine configured with three different IP addresses. On the left is the display of ProcessExplorer showing a process list that contains three different groups of duplicated IIS service processes. On the right are three IEs showing HTML and ASP pages returned from these three duplicated IIS web servers.

*4.1.3. Duplicating Other Services*

After duplicating RPCSS and Dcomlaunch, we further duplicated other services, including MySQL for Mysql database, Apache2.2 for Apache web server, Tlntsvr for telnet server, CiSvc for indexing files, ImapiService for managing CD recording, stisvc for providing image acquisition services for scanners and cameras, WebTool and WebClient for testing web server performance, sdCoreService and sdAuxService for protecting the system from malware and spyware, IIS service group, etc.

Figure 6 shows a screen shot of concurrently running three mysql database management server instances on a single Windows machine configured with three different IP addresses. On the left is the display of ProcessExplorer showing a process list that contains three groups of duplicated MySQL service's processes. On the right are three command-line windows displaying successful login results.

## 4.2 Performance Evaluation

Table 2

System call interception overhead: The Windows service duplication scheme only adds fewer than 0.1% extra CPU cycles. "Native" represents native Windows without FVM enforced, "New" and "Old" FVM represent FVM with and without service duplication respectively.

| System Calls | Native (CPU Cycles) | Old FVM (CPU Cycles) | New FVM (CPU Cycles) | Overhead |
| --- | --- | --- | --- | --- |
| NtCreateFile | 318398 | 382589(20%) | 382750(20%) | 0.04% |
| NtOpenFile | 159552 | 206458(29%) | 206625(30%) | 0.08% |
| NtQueryFullAttributesFile | 184507 | 223240(21%) | 223251(21%) | 0.01% |
| NtSetInformationFile | 38827 | 39992(3%) | 39995(3%) | 0.01% |
| NtOpenSemaphore | 29938 | 63654(113%) | 63858(113%) | 0.32% |
| NtCreatePort | 36885 | 71605(94%) | 71774(95%) | 0.24% |
| NtOpenSection | 37766 | 72045(91%) | 72131(91%) | 0.12% |

Table 3

Win32 API function interception overhead: The two FVM versions consume almost the same number of CPU cycles as compared with the baseline Windows. "Native" represents native Windows without FVM enforced, "New" and "Old" FVM represent FVM with and without service duplication respectively.

| Win32 API Functions | Native (CPU Cycles) | Old FVM (CPU Cycles) | NewFVM(CPU Cycles) | Overhead |
| --- | --- | --- | --- | --- |
| StartService | 2123732083 | 2123742943(<0.1%) | 2123742012(<0.1%) | <0.1% |
| RegisterServiceCtrlHandlerEx | 2808410 | 2808640(<0.1%) | 2808629(<0.1%) | <0.1% |
| QueryServiceStatusEx | 1971947 | 1971962(<0.1%) | 1971961(<0.1%) | <0.1% |
| SetServiceStatus | 2054881 | 2054903(<0.1%) | 2054906(<0.1%) | <0.1% |
| GetServiceDisplayName | 3565502 | 3565550(<0.1%) | 3565529(<0.1%) | <0.1% |
| CreateService | 8100322 | 8239648(<1.7%) | 8100649(<0.1%) | -1.7% |
| OpenService | 5381292 | 5381417(<0.1%) | 5478445(<1.8%) | 1.8% |

In this section, we examine the impact of Windows service duplication on the start-up latency and run-time performance of duplicated Windows services, such as the duplicated IIS server and Telnet server. The test-bed used in this evaluation consists of two machines. Machine A contains a Intel Celeron G4900 3.10GHz CPU with 8GB memory running Windows 7 and 10. Machine B contains an Intel Core 2 Duo 2GHz CPU with 4GB memory running Windows 7 and 10. We installed the old FVM (without service duplication) and the new FVM (with service duplication) on both machines.

As the performance overhead of service duplication mainly comes from additional instructions when intercepting system calls and API functions, we measured the interception overhead accordingly. First, we disabled the FVM layer, ran a group of services and applications natively in the host environment, and measured the average number of CPU cycles taken in each system call and API function through the *rdtsc* instruction. Then, we enabled the FVM layer without service duplication, ran the same services and applications in a container and took the same measurements. Finally, we enabled the FVM layer with service duplication, ran the same services and took the same measurements. Each of the reported numbers shown in Table 2 and 3 is an average of the results of 100 runs.

Table 2 shows a set of intercepted system calls and their average CPU cycles, including four file-related system calls and three IPC-related system calls. The FVM layer with service duplication consumes 3%~30% more CPU cycles than the native configuration for file-related system calls, and 91%~113% more CPU cycles for IPC-related system calls. Although the per-system call overhead appears to be significant for IPC-related system calls, the end-to-end impact on the overall system performance is much smaller, because IPC-related system calls account for lower than 0.2% of all invoked system calls in our test applications. Moreover, compared with the old FVM without service duplication, the current service duplication prototype only adds less than 0.3% extra CPU cycles. From these results, we can safely conclude that the performance cost of service duplication is quite acceptable.

Table 3 shows the average latency penalty in terms of CPU cycles for a set of intercepted Win32 API functions running on the baseline Windows OS, FVM without service duplication and FVM with service duplication. The two FVM versions consume almost the same number of CPU cycles as compared with the baseline Windows. Moreover, for most functions, the FVM version with service duplication is actually slightly faster than the FVM version without service duplication, because service names used in API functions do not need to be renamed when service duplication is in place. In addition, the higher performance differences (i.e., -1.7% and 1.8% respectively) associated with functions CreateService() and OpenService() are due to the extra communications between FVM modules at user level and kernel level.

Figure 7 shows the startup time of nine Windows services on machine B, including RPCSS, IISADMIN, and W3SVC running on Windows 7, and RPCSS, Dcomlaunch, and Tlntsvr running on Windows 10. Each of the test services was started in the following two ways: (a) starting the test service in the host; (b) simultaneously starting one, two and three instances of the test service respectively on FVM with service duplication. The startup time of a service corresponds to the elapsed time from the moment when the API function OpenSCManager() is called to the moment when the API function QueryServiceStatusEx() returns with the result code of SERVICE_RUNNING.

Figure 7 shows that the startup time of the last started instance of each test service on FVM when zero, one, two or three service instances are already active is almost identical to that of the same test service started in the same host. This suggests that the additional startup penalty due to FVM's service duplication is negligible, and remains so even when multiple service instances are started and duplicated simultaneously.

To assess the end-to-end performance impact of the proposed Windows service duplication scheme, we measured the performance of the IIS web server when multiple instances of it are duplicated and executed on a Windows 7 machine, and the performance of the Apache web server with the same set-up on a Windows 10 machine as well. Figure 8 shows the average throughputs of three IIS web server instances when they are duplicated and run on new FVM (with service duplication). The "Host" series shows the throughput of one IIS instance that runs in the host environment. Figure 9 shows similar results for three Apache web server instances. We measured the web servers' performance using WebBench [12], a licensed PC Magazine benchmark program, and each reported measurement is an average of results from three runs. Web servers were deployed on machine A, and configured in the same way when it runs on FVM and the host. The WebBench controllers and clients were deployed on machine B. In every testing session, each web server had one to twenty clients concurrently sending requests to it, and each client had only one outstanding request at a time. The results in Figure 8 and 9 show that, regardless of whether the underlying web server software is IIS or Apache, the average throughput of the three web server instances when they run on new FVM is about 5% lower than that of running in the host (i.e., native).

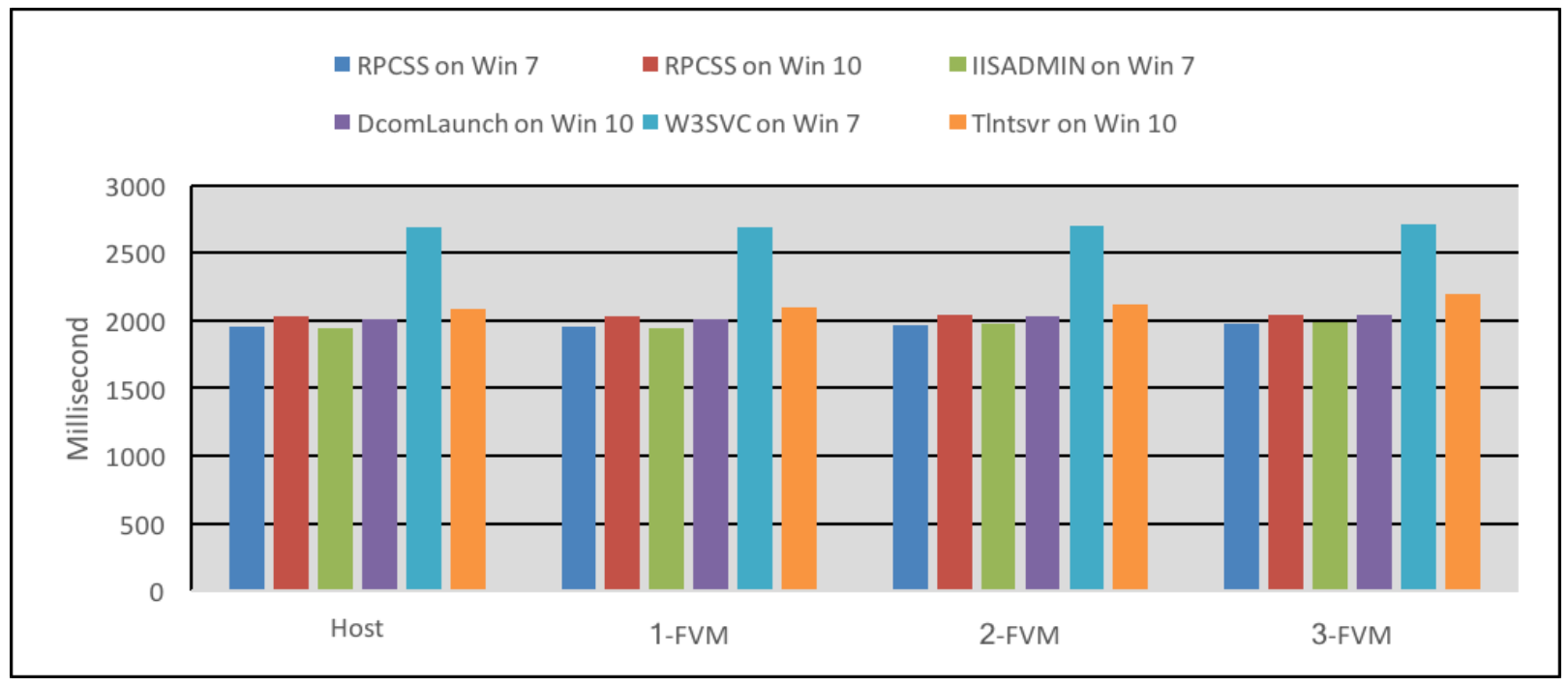

Fig. 7. Service startup time. Additional startup penalty of FVM's service duplication mechanism is negligible comparing to startup in the host.

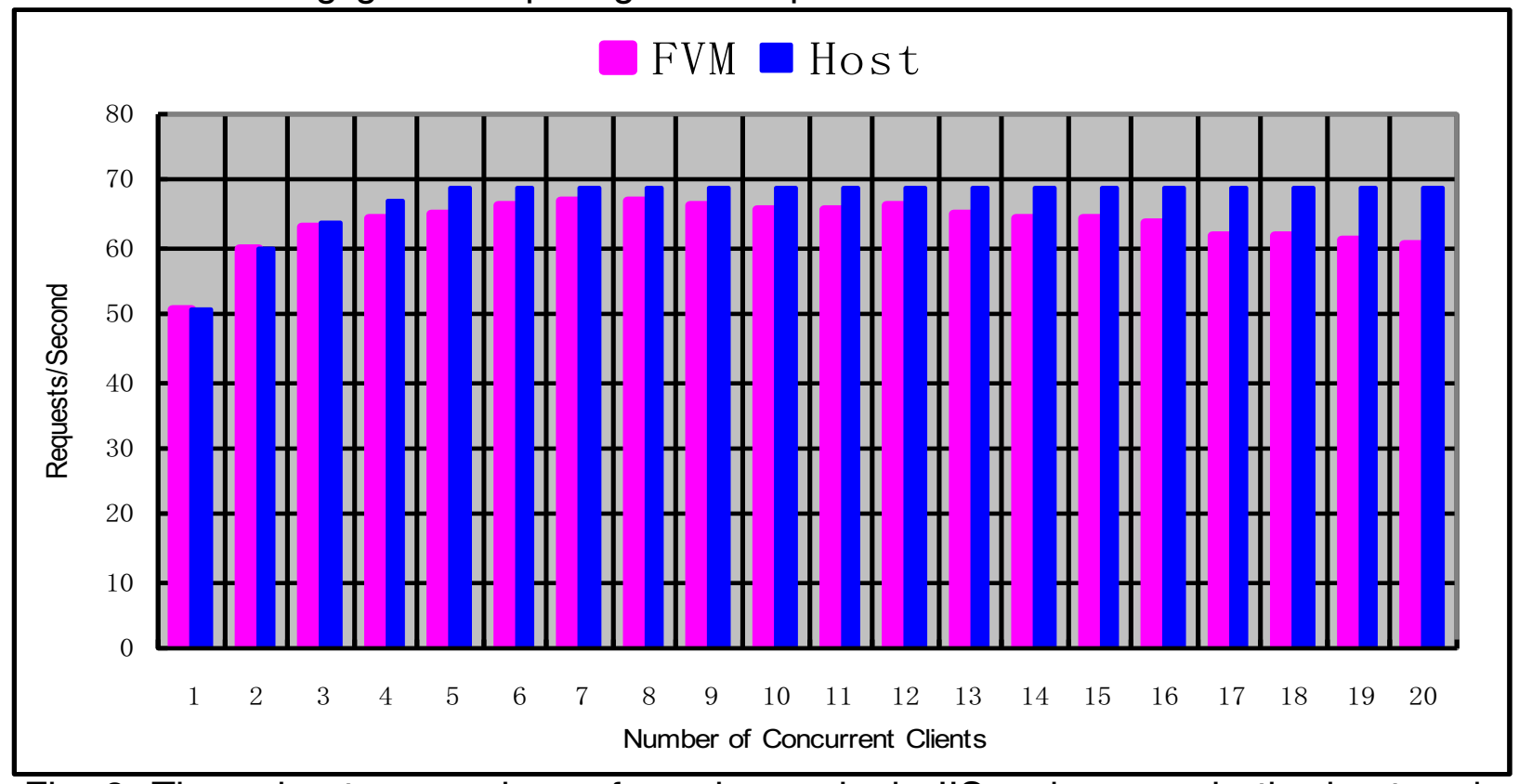

Fig. 8. Throughput comparison of running a single IIS web server in the host and three on the FVM with service duplication (i.e., new FVM). The average throughput of the three web servers on the FVM is slightly less than that of in the host.

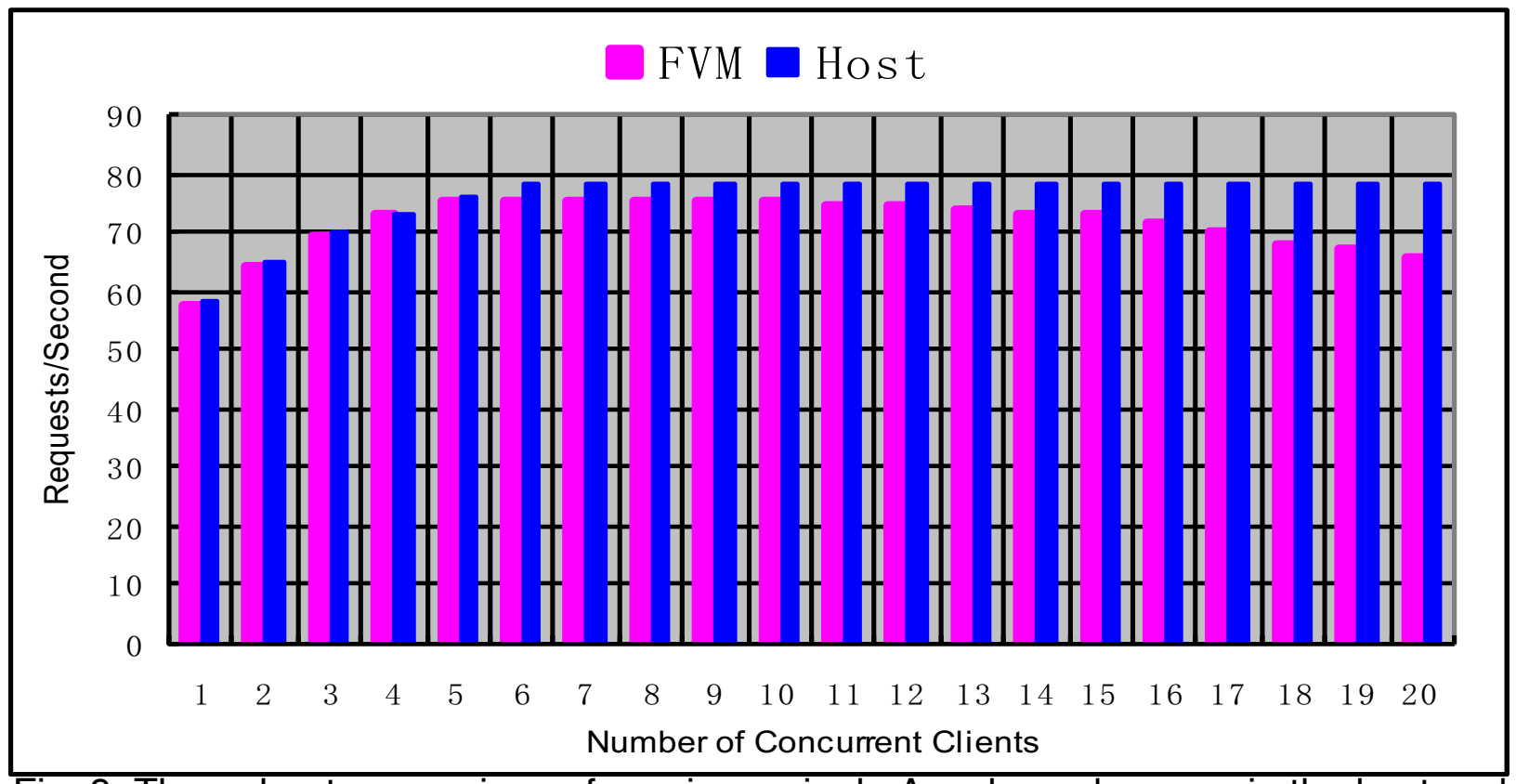

Fig. 9. Throughput comparison of running a single Apache web server in the host and three on the FVM with service duplication (i.e., new FVM). The average throughput of the three web servers on the FVM is slightly less than that of in the host.

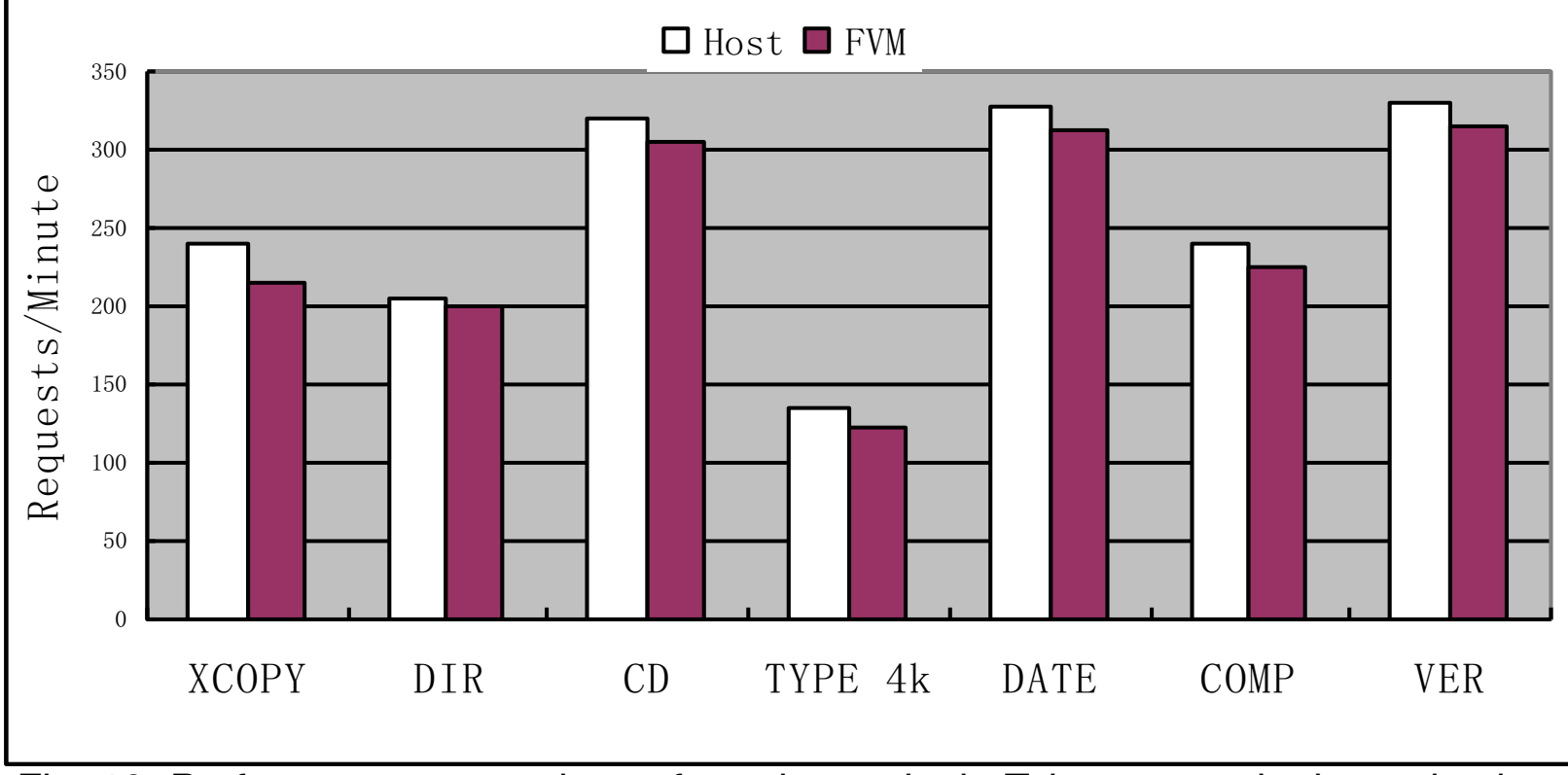

Fig. 10. Performance comparison of running a single Telnet server in the native host and three on the FVM with service duplication. The average command completion rate of FVM is slightly lower than the native host.

Figure 10 shows the performance of eight types of telnet commands when three Telnet server instances run on FVM and a single instance runs in the host. We used a program that can spawn multiple Telnet clients, which concurrently submit telnet commands and process responses from telnet servers. Each reported performance result is the average of the aggregate measurements on the three Telnet servers. The average performance of the three Telnet servers running on FVM is about 6% slower than that of the host.

In summary, the additional per-system-call or per-API-function-call overhead due to Windows service duplication is small to negligible when compared with the baseline Windows OS, and its penalty on the startup time and the run-time performance of duplicated services is also small (5%~6%). Moreover, compared with the old FVM, the new FVM introduce negligible extra reduction on the performance of system calls and API functions.

## 5 Related work

There are two projects more closely related to this work. One is Feather-weight Virtual Machine (FVM) [5], which enables multiple containers to run on a single Windows kernel in an isolated fashion. It is able to duplicate a limited number of ordinary services, but it can not duplicate system services such as RPCSS and ordinary services that require complex inter-service interactions, e.g., IIS service group. The other project is Virtuozzo [8] that provides isolated environments called Virtual Dedicated Server or Virtual Private Server (VPS) on Windows platform, but from its public documents, we could not find any description about Windows service duplication, especially system service duplication.

The original version of FVM can only duplicate a limited number of Windows services by intercepting service-related API functions and attaching FVM flags and container identification numbers to service names. However, as explained in Section 3.1, this method is not only limited but also incorrect in some cases because not all Windows services can be duplicated. For instance, the RPCSS service runs as a thread inside a service host process and therefore cannot be duplicated by just calling the standard service API function. Furthermore, because the original version of FVM did not recognize let alone address the "hard-coded service name" problem, it can not handle services whose binary contains hard-coded service names.

Other commercial products on Windows also include similar OS-level virtualization techniques, including Softricity Desktop [9], AppStream [10], PDS [20], and Thinstall [11]. In particular, Softricity Desktop [9] implements comprehensive virtualization to execute duplicated applications without requiring any pre-installation. Specifically it enables each duplicated application to execute against its own set of registries and configuration files within a container on any machine to which the duplicated application is deployed. However, Softricity Desktop can only provide isolated runtime environments for applications but not for Windows services.

Recent studies [16][34][35] build library OSes on host OSes, on which every application runs within its own address space with its own copy of the library OS. The library of a Windows service is copied into the address space of an application, and thus library OS can run multiple instances of a Windows service simultaneously inside multiple applications respectively. However, the library OS is significantly different from OS-level virtualization. A container can support multiple applications and the interactions among them, but a copy of library OS can only support the host application. More importantly, the library OS is difficult to support multi-process applications which are commonly used and require communications through shared states in Windows system libraries.

On Linux OS, there are also several successful OS-level virtualization systems, such as Zap [17], Linux-VServer [13], and OpenVZ [18]. Zap provides a virtualization mechanism to form pods to transparently migrate legacy and networked applications across independent machines. A pod is a group of processes that are provided a consistent, virtualized view of the system. Linux-VServer and OpenVZ are container-based systems that provide a shared, virtualized OS image consisting of a root file system, a set of system libraries and executables. Each container can be booted, shut down, and rebooted just like a regular operating system. However, in this work, we do not try to design just another type of OS-level virtualization mechanism. Instead, we aim to solve an important problem on various OS-level virtualization platforms, that is, duplicating services on Windows or daemons on Linux. Existing projects did not systematically resolve all of the issues related to the service duplication, as analyzed in Section 3.1. For example, they are not aware of and thus can not address the issues of hard-coded service names and DLL-based services.

## 6 Conclusion

OS-level virtualization virtualizes system resources at the system call interface, and relies on a single kernel to isolate multiple containers running in parallel on top of it via resource renaming. On the Windows platform, additional complications arise because parts of the kernel's functionalities are actually embodied in some user-level services, which are tightly integrated with the OS and therefore cannot be easily duplicated or virtualized in each container. Lack of effective solutions to duplicate these Windows services has been the Achilles' heel for OS-level virtualization to become a viable virtualization technology on Windows. Unfortunately, none of the publicly available documents on OS-level virtualization technologies ever mention the service duplication problems, not to mention addressing them.

This paper presents the first known Windows service duplication methodology that can duplicate not only ordinary Windows services, but also several critical Windows system services such as RPCSS. By applying this proposed methodology, we have successfully duplicated RPCSS, DcomLaunch, and the IIS service group including IISADMIN, W3SVC, MSFTPSVC, SMTPSVC, and NNTPSVC, as well as MySQL, Apache2.2, CiSvc, ImapiService, and Tlntsvr on different Windows platforms. As a result, we can run multiple instances of the IIS web server, Apache web server or MySQL database server simultaneously on a single Windows machine, a feat that no previous OS-level virtualization systems on the Windows platform, including commercial offerings, is able to accomplish.

Moreover, we also develop a useful technique to facilitate the communications between an application in a container and a process in the host environment. We address the issue by employing a hash table to automatically identify the IPC objects that are conducting the communications across container boundaries.

Detailed performance measurements show that the additional performance overhead introduced by the proposed Windows service duplication scheme is rather minor when compared with the overhead introduced by the baseline OS-level virtualization logic. Furthermore, the startup time and run-time performance of duplicated services of FVM with service duplication are slightly lower than the same set-ups running in the host. These results effectively demonstrate the potential performance advantage and thus the overall effectiveness of the proposed Windows service duplication technique.

## ACKNOWLEDGMENT

We would like to thank all the anonymous reviewers for their insightful comments and feedback. This work is supported by US Army Research Lab under grant "Models for Enabling Continuous Reconfigurability of Secure Missions".